\documentclass{emulateapj}

\newcommand{\ee}[1]{\!\times\!10^{#1}}
\newcommand{\gws}{gravitational waves~}
\newcommand{\gw}{gravitational wave~}

\begin{document}

\title{Beating the spin-down limit on gravitational wave emission from the Crab
pulsar}


\author{B.~Abbott\altaffilmark{16},
R.~Abbott\altaffilmark{16},
R.~Adhikari\altaffilmark{16},
P.~Ajith\altaffilmark{2},
B.~Allen\altaffilmark{2, 55},
G.~Allen\altaffilmark{33},
R.~Amin\altaffilmark{20},
S.~B.~Anderson\altaffilmark{16},
W.~G.~Anderson\altaffilmark{55},
M.~A.~Arain\altaffilmark{42}
% ...
}

\affil{The LIGO Scientific Collaboration, http://www.ligo.org}

\author{
G.~Santostasi\altaffilmark{23}
}


\altaffiltext {1}{Albert-Einstein-Institut, Max-Planck-Institut f\"ur Gravitationsphysik, D-14476 Golm, Germany}
\altaffiltext {2}{Albert-Einstein-Institut, Max-Planck-Institut f\"ur Gravitationsphysik, D-30167 Hannover, Germany}
\altaffiltext {3}{Andrews University, Berrien Springs, MI 49104 USA}
\altaffiltext {4}{Australian National University, Canberra, 0200, Australia}
\altaffiltext {5}{California Institute of Technology, Pasadena, CA  91125, USA}
\altaffiltext {6}{Caltech-CaRT, Pasadena, CA  91125, USA}
\altaffiltext {7}{Cardiff University, Cardiff, CF24 3AA, United Kingdom}
\altaffiltext {8}{Carleton College, Northfield, MN  55057, USA}
\altaffiltext {9}{Charles Sturt University, Wagga Wagga, NSW 2678, Australia}
\altaffiltext {10}{Columbia University, New York, NY  10027, USA}
\altaffiltext {11}{Embry-Riddle Aeronautical University, Prescott, AZ   86301 USA}
\altaffiltext {12}{Hobart and William Smith Colleges, Geneva, NY  14456, USA}
\altaffiltext {13}{Institute of Applied Physics, Nizhny Novgorod, 603950, Russia}
\altaffiltext {14}{Inter-University Centre for Astronomy  and Astrophysics, Pune - 411007, India}
\altaffiltext {15}{Leibniz Universit{\"a}t Hannover, D-30167 Hannover, Germany}
\altaffiltext {16}{LIGO - California Institute of Technology, Pasadena, CA  91125, USA}
\altaffiltext {17}{LIGO - Massachusetts Institute of Technology, Cambridge, MA 02139, USA}
\altaffiltext {18}{LIGO Hanford Observatory, Richland, WA  99352, USA}
\altaffiltext {19}{LIGO Livingston Observatory, Livingston, LA  70754, USA}
\altaffiltext {20}{Louisiana State University, Baton Rouge, LA  70803, USA}
\altaffiltext {21}{Louisiana Tech University, Ruston, LA  71272, USA}
\altaffiltext {22}{Loyola University, New Orleans, LA 70118, USA}
\altaffiltext {23}{McNeese State University, Lake Charles, LA 70609, USA}
\altaffiltext {24}{Moscow State University, Moscow, 119992, Russia}
\altaffiltext {25}{NASA/Goddard Space Flight Center, Greenbelt, MD  20771, USA}
\altaffiltext {26}{National Astronomical Observatory of Japan, Tokyo  181-8588, Japan}
\altaffiltext {27}{Northwestern University, Evanston, IL  60208, USA}
\altaffiltext {28}{Rutherford Appleton Laboratory, Chilton, Didcot, Oxon OX11 0QX United Kingdom}
\altaffiltext {29}{San Jose State University, San Jose, CA 95192, USA}
\altaffiltext {30}{Sonoma State University, Rohnert Park, CA 94928, USA}
\altaffiltext {31}{Southeastern Louisiana University, Hammond, LA  70402, USA}
\altaffiltext {32}{Southern University and A\&M College, Baton Rouge, LA  70813, USA}
\altaffiltext {33}{Stanford University, Stanford, CA  94305, USA}
\altaffiltext {34}{Syracuse University, Syracuse, NY  13244, USA}
\altaffiltext {35}{The Pennsylvania State University, University Park, PA  16802, USA}
\altaffiltext {36}{The University of Texas at Austin, Austin, TX 78712, USA}
\altaffiltext {37}{The University of Texas at Brownsville and Texas Southmost College, Brownsville, TX  78520, USA}
\altaffiltext {38}{Trinity University, San Antonio, TX  78212, USA}
\altaffiltext {39}{Universitat de les Illes Balears, E-07122 Palma de Mallorca, Spain}
\altaffiltext {40}{University of Adelaide, Adelaide, SA 5005, Australia}
\altaffiltext {41}{University of Birmingham, Birmingham, B15 2TT, United Kingdom}
\altaffiltext {42}{University of Florida, Gainesville, FL  32611, USA}
\altaffiltext {43}{University of Glasgow, Glasgow, G12 8QQ, United Kingdom}
\altaffiltext {44}{University of Maryland, College Park, MD 20742 USA}
\altaffiltext {45}{University of Massachusetts, Amherst, MA 01003 USA}
\altaffiltext {46}{University of Michigan, Ann Arbor, MI  48109, USA}
\altaffiltext {47}{University of Minnesota, Minneapolis, MN 55455, USA}
\altaffiltext {48}{University of Oregon, Eugene, OR  97403, USA}
\altaffiltext {49}{University of Rochester, Rochester, NY  14627, USA}
\altaffiltext {50}{University of Salerno, 84084 Fisciano (Salerno), Italy}
\altaffiltext {51}{University of Sannio at Benevento, I-82100 Benevento, Italy}
\altaffiltext {52}{University of Southampton, Southampton, SO17 1BJ, United Kingdom}
\altaffiltext {53}{University of Strathclyde, Glasgow, G1 1XQ, United Kingdom}
\altaffiltext {54}{University of Western Australia, Crawley, WA 6009, Australia}
\altaffiltext {55}{University of Wisconsin-Milwaukee, Milwaukee, WI  53201, USA}
\altaffiltext {56}{Washington State University, Pullman, WA 99164, USA}


\shorttitle{Beating the Crab pulsar spin-down limit}
\shortauthors{The LIGO Scientific Collaboration}

\begin{abstract}
We present direct upper limits on gravitational wave emission from the Crab
pulsar using data from the first nine months of the fifth science run of the
Laser Interferometer Gravitational-wave Observatory (LIGO). These limits are
based on two searches. In the first we assume that the gravitational wave
emission follows the observed radio timing, giving an upper limit on
gravitational wave emission that beats indirect limits inferred from the
spin-down and braking index of the pulsar and the energetics of the nebula. In
the second we allow for a small mismatch between the gravitational and radio
signal frequencies and interpret our results in the context of two possible
gravitational wave emission mechanisms.
\end{abstract}

\section{Introduction}\label{sec:intro}
The Crab pulsar (PSR\,B0531$+$21, PSR\,J0534$+$2200) has long been regarded as
one of the most promising \emph{known} local sources of \gw emission and is an
iconic target for gravitational wave searches \citep{PressThorne:1972,
Zimmermann:1978}. Its high spin-down rate, $\dot{\nu} \approx -3.7\ee{-10}\,{\rm
Hz}\,{\rm s}^{-1}$, corresponds to a kinetic energy loss rate of $\dot{E} =
4\pi^2I_{zz}\nu|\dot{\nu}| \approx 4.4\ee{31}$\,W (using a spin frequency of
$\nu = 29.78$\,Hz and the canonical value of $10^{38}$\,kg\,m$^2$ for the
principal moment of inertia $I_{zz}$.) This loss is due to a variety of
mechanisms, including magnetic dipole radiation, particle acceleration in the
magnetosphere, and gravitational radiation. If one assumes that all the energy
\emph{is} being radiated gravitationally, the \gw tensor amplitude at Earth is
$h_0^{\rm sd} = 8.06\ee{-19}\,I_{38}r_{\rm kpc}^{-1}(|\dot{\nu}|/\nu)^{1/2}$,
where $r_{\rm kpc}$ is the distance to the pulsar in kpc and $I_{38}$ is the
moment of inerta in units of the canonical value \citep{ab3}. For the Crab
pulsar this ``spin-down upper limit'' is $h_0^{\rm sd} = 1.4\ee{-24}$, using the
canonical moment of inertia and a distance $r=2$\,kpc. It has long been known
that the Laser Interferometer Gravitational-wave Observatory (LIGO) can achieve
this sensitivity by integrating several months of data with the initial design
noise spectrum.

The electromagnetic emission and accelerating expansion of the Crab Nebula are
powered almost entirely by the rotation of the pulsar. The question now is
whether these two loss mechanisms can account for the vast majority of the
observed rotational energy loss, or whether \gw emission has a significant part
to play.

The bolometric luminosity of the nebula is (1--2)$\ee{31}$~W, which accounts for
less than half the spin-down power \citep[e.g.,][]{DavidsonFesen:1985}. There
have been many attempts to estimate the power involved in the observed
acceleration of optical filaments, for example recently by \citet{Bejger:2002,
Bejger:2003}. However these depend on poorly known factors such as the mass and
expansion history of the nebula, and the uncertainties in the estimated power
are comparable to the spin-down power itself. Thus electromagnetic observations
of the nebula, within their uncertainties, still allow for a substantial
fraction of the spin-down power to be emitted in gravitational waves.

The braking index $n=\nu \ddot{\nu} /\dot{\nu}^2$ of the pulsar further
constrains the gravitational wave emission. The observed value $n=2.5$ still is
not well understood on theoretical grounds, but since quadrupolar radiation has
$n=5$ it implies that only a small fraction of the spin-down power is emitted in
gravitational waves. The best estimate in print is by \citet{Palomba:2000}, who
uses a phenomenological model of the spin-down (present and historical) together
with the present braking index and known age of the pulsar to estimate that the
highest possible $h_0$ today is about 40\% of the spin-down limit. This value is
consistent with the observations of the nebula, and is also observable with
several months of data from LIGO's fifth science run (S5).

Early directed searches for \gws from the Crab pulsar were performed by
\citet{Levine:1972}, using a 30\,m interferometer to give a strain upper limit
of $3\ee{-17}$, and \citet{Hirakawa:1978}, using a bar detector. The most recent
bar result \citep{Suzuki:1995} gave an upper limit that was still over an order
of magnitude above the spin-down limit. The LIGO detectors have improved on
these results, with LIGO's second science run (S2) producing a 95\% upper limit
of $h_0^{95\%} = 4.1\ee{-23}$ \citep{ab2}, and the combined data from the S3 and
S4 runs produced an upper limit of $h_0^{95\%} = 3.1\ee{-24}$ \citep{ab3} only
2.2 times greater than the spin-down limit.

In this Letter, we describe searches of data from the fifth LIGO science run,
which started on 2005 November 4 and ended on 2007 October 1 \citep{Abbott:S5}.
During this period the detectors (the 4\,km and 2\,km detectors at LIGO Hanford
Observatory, H1 and H2, and the 4\,km detector at the LIGO Livingston
Observatory, L1) were at their design sensitivities and had duty factors of
$78\%$ for H1, $79\%$ for H2, and $\sim66\%$ for L1. The GEO600 detector
\citep{Luck:2006} also participated in the S5 run but was much less
sensitive at the frequency of the expected signal.

The Crab pulsar was observed to glitch on 2006 August 23 at approximately 04:00
UTC \citep{CrabEphemeris, Lyne:PC}. Since the glitch mechanism is not certain
and may involve unpredictable changes in the gravitational wave timing
and amplitude, we use this glitch as natural point at which to pause this
coherent search for the Crab pulsar. Our data set consists of H1 and H2 data
from 2005 November 4 and L1 data from 2005 November 14 up to 2006 August 23. For
the two different searches carried out in this analysis, described below, this
gives 201, 222 and 158 days of data for H1, H2 and L1 respectively for the
single-template search, and 182, 206, and 141 days of data for the
multi-template frequency-frequency first derivative search, which required
larger contiguous segments than the single-template search.

\section{Methods}
We use two different methods (see \citealt{ab1}) to search for \gws from the
Crab pulsar to account for different emission scenarios. One method uses a
single time domain template for the \gw signal assuming that the \gw period
evolves precisely as the electromagnetic pulse period. The other method works in
the frequency domain to cover a relatively small, physically motivated range of
frequency and spin-down values. The searches use the known frequency and
position of the Crab pulsar, as derived from the Jodrell Bank Crab Pulsar
Monthly Ephemeris \citep{CrabEphemeris}. Using this ephemeris and the
assumption that the \gw and electromagnetic phase track each other
precisely, we can predict the signal phase evolution with negligible uncertainty.
Both searches assume that emission will be at or near twice the pulsar's spin
frequency, $2\nu = \nu_{\rm GW}\sim59.56$\,Hz, which is the frequency of
emission by a steadily rotating quadrupolar deformation, i.e.\ a triaxial star.
The Crab pulsar might be emitting at $\nu_{\rm GW} \approx 4\nu/3$ through an
$r$-mode \citep{Owen:1998} if the mode saturates at a small amplitude and thus
is long-lived \citep[e.g.,][]{Brink:2004}. However, the uncertainty of this
frequency is relatively large, of order one part in $10^3$
\citep{Lindblom:1999}. Due to this, and the greater instrument noise at this
frequency, we did not search for $r$-modes. Although $2\nu$ is close to
the 60\,Hz power line frequency, it is sufficiently far away that the searches
are relatively unaffected by non-stationary components of the power line noise.
The absolute timing accuracy of the LIGO data is sufficiently good that the
likelihoods produced for each detector can be combined to give a joint
likelihood.

For a given search frequency and spin-down, the four unknown signal parameters
are the \gw amplitude $h_0$, the initial phase $\phi_0$, the spin-axis
inclination angle $\iota$, and the polarization angle $\psi$. X-ray
 observations of the Crab Pulsar Wind Nebula provide values of the orientation
angle $\iota$ and polarisation angle $\psi$ of the pulsar. From \citet{Ng:2004,
Ng:2008} we use $\iota = 62.17\pm2.195^{\circ}$ and $\psi =
125.155\pm1.355^{\circ}$, where we have taken the mean of the best fit values
for the outer and inner tori of the nebula. We use these ranges to put Gaussian
priors on these two parameters for both the search techniques. On the chance
that the star is misaligned from these structures, we also present results using
uniform priors over the allowed ranges of the parameters.

The single-template search \citep{DupuisWoan:2005} assumes a triaxial star
emitting \gws at precisely twice the spin frequency, following the
electromagnetic pulse phase evolution and taking into account the small
variations in phase caused by timing noise \citep{PitkinWoan:2004}. It uses a
standard Bayesian methodology to produce a joint posterior probability volume
over the four unknown parameters using data from all three detectors. We
use both uniform priors and restricted priors on $\psi$ and $\iota$ when
calculating the posterior. We marginalize the angle parameters to produce a
posterior probability for $h_0$ and from this calculate a 95\%
degree-of-belief upper limit on the \gw amplitude.

A search was also performed at \gw frequencies $\nu_{\rm GW}$ in a narrow
band about $2\nu$, based on simple astrophysical arguments. We begin by writing
$\nu_{\rm GW} = 2 \nu (1+\delta)$, where $\delta$ is a small number.  A relation
of this form holds if the \gws are produced by a component spinning separately
from the electromagnetically emitting one, with the two components linked by
some torque which acts to enforce co-rotation between them on a timescale
$\tau_{\rm coupling}$.  In such a case $ \delta \sim \tau_{\rm coupling} /
\tau_{\rm spin-down}$, where $\tau_{\rm spin-down} \sim \nu / \dot \nu \simeq
2500$ years. A relation of the form given for $\nu_{\rm GW}$ above also holds if
the \gws are produced by free precession of a nearly biaxial star
\citep{JA:2002}. In such a case $\delta \sim \alpha (I_{zz}-I_{xx})/I_{xx}$
where $\alpha$ is a factor of order unity dependent on the geometry of the
free precession, e.g.\ the angle between the symmetry axis and angular momentum
axis. No clear signature of free precession has been seen in the radio
pulsations of the Crab pulsar, although precession would have little effect on
the radio signal if the amplitude of the precession were small.

Together, these scenarios suggest searching over a frequency interval $\pm
\Delta \nu_{\rm GW}$ centred on $2\nu$, where $\Delta\nu_{\rm GW} \sim |\delta|
\, 2\nu$.  We have followed such a strategy, using a maximum value of
$|\delta| = 10^{-4}$. In terms of the two-component model, such a $|\delta|$
value corresponds to $\tau_{\rm coupling} \sim 10^{-4} \, \tau_{\rm spin-down}
\sim$ several months, comparable to the longest timescales seen in glitch
recovery where re-coupling between the two components might be expected to
occur. In terms of free precession, $|\delta| = 10^{-4}$ is on the high end of
the range of deformations that compact objects are thought to be capable of
sustaining \citep{Owen:2005, Lin:2007, Haskell:2007}.

Using the above estimates as a guide, a band of frequencies $\pm
6\times10^{-3}$\,Hz centred on twice the Crab pulsar's observed frequency was
searched over. Corresponding bands in frequency derivatives were motivated via
differentiation of the equation for $\nu_{\rm GW}$, which together with the
assumption that $\delta$ itself evolves no more rapidly than on the spin-down
timescale, leads to a band in frequency first derivative of $\pm
1.5\times10^{-13}$ Hz/s, with searches over higher derivatives being
unnecessary.

The multi-template search method is a maximum likelihood technique, the coherent
multi-detector $F$-statistic derived in \citet{CutlerSchutz:2005}. An explicit
search is required over a single sky position and second derivative of the
frequency, and over the selected ranges of the frequency and of the first
frequency derivative. The spacing of the templates is chosen in such a way as to
ensure at most a 5\% loss in the detection statistic, resulting in a total of
$3\times10^{7}$ templates. The detection statistic $2F$ is computed for each
template. The expected $3\sigma$ range of the largest $2F$ value for Gaussian
noise (no signal present) and $3\times10^{7}$ templates is 35--49. The largest
$2F$ value found in the actual search is 37, well within the expected range for
noise.

Based on the largest $2F$ value, 95\% confidence upper limits are produced using
a frequentist Monte Carlo injection method, as described in \citet{ab7}. For the
unknown parameters uniform distributions and physically informed distributions
were used for the injected population of signals, consistent with the choices
made for the single-template time domain search.

\section{Results}\label{sec:results}
In the single-template search the joint (i.e.\ multi-detector) posterior
probability distribution for the \gw amplitude peaks at zero, indicating that no
signal is visible at our current sensitivity. The joint 95\% upper limit on the
\gw amplitude, using uniform priors on all the parameters, is $h_0^{95\%} =
3.4\ee{-25}$. In terms of the pulsar's ellipticity, given by $\varepsilon =
0.237\,h_{-24}r_{\rm kpc}\nu^{-2}I_{38}$ \citep{ab3}, where $h_{-24}$ is $h_0$
in units of $1\ee{-24}$, this gives $\varepsilon = 1.8\ee{-4}$ using the
canonical moment of inertia and $r=2$\,kpc. This is 4.1 times lower than the
spin-down upper limit and also 1.6 times lower than the limit estimated by
\citet{Palomba:2000} (see \S\ref{sec:intro}.) Squaring the ratio of the
spin-down and direct upper limit shows that less than $\approx6$\% of the total
power available from spin-down is being emitted as gravitational waves, assuming
the canonical moment of inertia. Using the restricted priors on $\psi$ and
$\iota$ we get an upper limit on $h_0$ of $2.7\ee{-25}$, which is 1.3 times
smaller than that with uniform priors, and corresponds to less than 4\% of the
spin-down energy available.

With the coherent multi-template frequency-frequency first derivative search we
set $95\%$ confidence upper limits on $h_0$ and ellipticity of $1.7\ee{-24}$
and $9.0\ee{-4}$ respectively, over the entire parameter space searched.
These upper limits are larger than the single-template search limits by
roughly a factor of five. This is to be expected because the larger number of
templates raises the number of trials and thus the statistical confidence
threshold. Assuming restricted priors on $\psi$ and $\iota$ yields an improved
upper limit of $1.2\ee{-24}$, a factor of 1.2 below the spin-down limit, across
the entire parameter space searched. This limits the energy budget of \gws to be
less than 73\% of the available energy. These quoted upper limits are subject to
uncertainty in the calibration of the detectors. Amplitude calibration
uncertainties for H1, H2 and L1, respectively, are: 8.1\%, 7.2\% and 6.0\%
(single-template analysis), and 9.5\%, 7.8\% and 8.7\% (multi-template
analysis).

\section{Discussion}\label{sec:discussion}
Under the assumption that the \gw and the electromagnetic signals are
phase-locked, our single-template search results constrain the gravitational
wave luminosity to be less than 6\% of the observed spin-down luminosity.
This beats the indirect limits inferred from all electromagnetic
observations of the Crab pulsar and nebula.

Our upper limits are interesting because they have entered the outskirts of the
range of theoretical predictions. Normal neutron stars are believed to be
mostly fluid with maximum elastic deformations orders of magnitude smaller than
the few $\times10^{-4}$ of our upper limits, but some theories of quark matter
predict solid or mostly solid stars which could sustain such ellipticities
\citep{Owen:2005, Lin:2007, Haskell:2007}. However, our upper limits do not
constrain the composition of the star and cannot constrain any fundamental
properties of quark matter. The ellipticity is proportional to the quadrupolar
strain, which may simply be very low for a given star no matter its composition.
The Crab is likely to have an ellipticity at least about $10^{-11}$ due to the
stresses of its internal magnetic field \citep{Cutler:2002} if the internal
field is comparable to the external dipole of $4\times10^{12}$~G. Our upper
limits can be interpreted as direct upper limits of about $10^{16}$~G on the
internal magnetic field, depending on the ratio of toroidal to poloidal
components \citep{Colaiuda:2007}.

\begin{figure}
\includegraphics[width=0.4\textwidth]{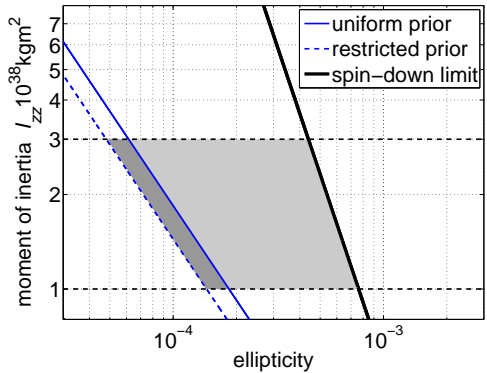}
\caption{The single template search upper limits from S5, for the uniform and
restricted prior ranges, and spin-down upper limit plotted as exclusion regions
in a moment of inertia--ellipticity plane. Areas to the right of the diagonal
lines are excluded. The dashed horizontal lines represent estimates of the
theoretical lower and upper bounds of acceptable moments of inertia at
($1$--$3$)$\ee{38}\,$kg\,m$^2$. The shaded area represents the region that is
newly excluded with these results.\label{fig:ieplane}}
\end{figure}

As discussed in \citet{ab3} there is considerable uncertainty in the true value
of the Crab pulsar's moment of inertia. The best guesses at its value come from
neutron star equation of state models rather than direct measurements. Previous
pulsar ellipticity upper limits and spin-down limits have made use of the
canonical value of $I_{zz}$. We can however cast our upper limit in a way that
makes no assumptions about the moment of inertia, by placing the limit on the
neutron star quadrupole moment $\approx I_{zz}\varepsilon$. This then allows us
to plot the single-template search results as exclusion regions in the
$I\textrm{-}\varepsilon$ plane. The results, with uniform and restricted prior
ranges, are plotted in this way in Figure~\ref{fig:ieplane}. Our upper limits
are smaller than the spin-down limit by a factor that varies as $I_{zz}^{1/2}$.
If we take the theoretical upper bound on the moment of inertia to be
$3\ee{38}$\,kg\,${\rm m}^2$ as in \citep{ab3} then the result with uniform
priors beats the spin-down limit by a factor of 7.2.

Finally, the physical interpretation of our multi-template search depends upon
the assumed cause of the splitting $\nu_{\rm GW} = 2\nu (1+\delta)$ between
gravitational and electromagnetic signals. In the context of the two-component
spin-down model, our results show that a gravitational wave emitting component
of the star coupled to the electromagnetic (radio) emitting component on a
timescale of a few months or less has a quadrupole asymmetry $I_{yy}-I_{xx}$ of
no more than $9.0 \times 10^{34}$\,kg\,m$^2$. This is about five times larger
than the bound on $I_{yy}-I_{xx}$ obtained in the single-template search. If
free precession is responsible for the frequency splitting our results instead
give an upper limit on the product $\Delta I \sin^2\theta$, where $\Delta I$ is
the $I_{zz}-I_{xx}$ part of the quadrupole moment tensor that participates in
the precession and $\theta$ the wobble angle \citep{JA:2002}.

\acknowledgements
The authors gratefully acknowledge the support of the United States
National Science Foundation for the construction and operation of the
LIGO Laboratory and the Science and Technology Facilities Council of the
United Kingdom, the Max-Planck-Society, and the State of
Niedersachsen/Germany for support of the construction and operation of
the GEO600 detector. The authors also gratefully acknowledge the support
of the research by these agencies and by the Australian Research Council,
the Council of Scientific and Industrial Research of India, the Istituto
Nazionale di Fisica Nucleare of Italy, the Spanish Ministerio de
Educaci\'on y Ciencia, the Conselleria d'Economia, Hisenda i Innovaci\'o of
the Govern de les Illes Balears, the Scottish Funding Council, the
Scottish Universities Physics Alliance, The National Aeronautics and
Space Administration, the Carnegie Trust, the Leverhulme Trust, the David
and Lucile Packard Foundation, the Research Corporation, and the Alfred
P. Sloan Foundation. LIGO Document No. LIGO-P070118-00-Z.

\end{document}